\documentclass[%
 reprint,
 amsmath,amssymb,
 aps,
]{revtex4-2}

\usepackage[colorlinks, citecolor=blue]{hyperref}
\usepackage{graphicx}
\usepackage{dcolumn}
\usepackage{bm}

\begin{document}

\title{ Sensitivity improvement of Rydberg atom-based microwave sensing via
electromagnetically induced transparency}\thanks{E-mail: liuhongping@wipm.ac.cn}
\author{M. H. Cai}
\affiliation{State Key Laboratory of Magnetic Resonance and Atomic and Molecular Physics,
Wuhan Institute of Physics and Mathematics, Innovation Academy for Precision
Measurement Science and Technology, Chinese Academy of Sciences, Wuhan 430071, China}
\affiliation{University of Chinese Academy of Sciences, Beijing 100049, China}

\author{Z. S. Xu}
\affiliation{State Key Laboratory of Magnetic Resonance and Atomic and Molecular Physics,
Wuhan Institute of Physics and Mathematics, Innovation Academy for Precision
Measurement Science and Technology, Chinese Academy of Sciences, Wuhan 430071, China}
\affiliation{University of Chinese Academy of Sciences, Beijing 100049, China}

\author{S. H. You}
\affiliation{State Key Laboratory of Magnetic Resonance and Atomic and Molecular Physics,
Wuhan Institute of Physics and Mathematics, Innovation Academy for Precision
Measurement Science and Technology, Chinese Academy of Sciences, Wuhan 430071, China}
\affiliation{University of Chinese Academy of Sciences, Beijing 100049, China}

\author{H. P. Liu}
\affiliation{State Key Laboratory of Magnetic Resonance and Atomic and Molecular Physics,
Wuhan Institute of Physics and Mathematics, Innovation Academy for Precision
Measurement Science and Technology, Chinese Academy of Sciences, Wuhan 430071, China}
\affiliation{University of Chinese Academy of Sciences, Beijing 100049, China}

\begin{abstract}
A highly excited Rydberg atom via electromagnetically induced transparency
with two color cascading lasers has extreme sensitivity to electric fields
of microwave ranging from 100 MHz to over 1 THz.
It can be
used as susceptible atom-based microwave communication antennas where the
carrier wave usually works exactly resonant to the transition between a pair
of adjacent Rydberg states with large electric dipole moment.
A technique of
superheterodyne with a strong on-resonant local microwave oscillator is
employed to induce considerable Autler-Townes splitting where the antennas
has a highest dynamic response to another weak target signal microwave
carrier.
To further improve the sensitivity of atomic antenna in
communication, we detune the carrier microwave frequency off resonance
forming an asymmetrically optical splitting and fix the coupling laser
frequency at the shoulder of the stronger one, and optimize the local field
strength simultaneously.
It gives a sensitivity of 12.50(04) $\rm{nVcm^{-1}\cdot Hz^{-1/2}}$.
Its enhancement mechanism of sensitivity is also proved by a
theoretical simulation.
\end{abstract}

\date{\today }
\maketitle

\volumeyear{ } \volumenumber{ } \issuenumber{ } \eid{ } \received[Received
text]{}

\revised[Revised text]{}

\accepted[Accepted text]{}

\published[Published text]{}

\startpage{1} \endpage{ }


\section{INTRODUCTION}

Rydberg atoms in highly excited states with one or more electrons of large
principal quantum numbers are sensitive to electric fields, very suitable to
manufacture atom-based sensor for detecting and receiving communication
signals \cite{a1}. It has been widely investigated thoroughly both
theoretically and experimentally throughout the last decades \cite{a206,
a182, a174, a195, a215, a221,89260,a192,92167,92735,91120,89261}. This type of sensors can replace the
front-end components and electronics in a conventional antenna/receiver
system \cite{a190, a214}, since they have potential advantages over
conventional systems. This Rydberg-atom based quantum sensor owns unique
properties such as self-calibration and fine spatial resolution in both the
far-field and near-field \cite{a203, a199, a226, a190}.

The spacing between Rydberg levels can locate across the microwave radio
frequency, which can be used to measure RF E-field strengths over a large
range of frequencies (1 GHz to 500 GHz) with a high sensitivity \cite{a186,
a195, a205}, approximately $<1\ \rm{{\mu Vcm^{-1}\cdot Hz^{-1/2}}}$ \cite{a183, a188}.
Various groups have investigated wireless communication using
Rydberg atoms \cite{a222, a206, a216, a219,a221}. For example, utilizing the
Rydberg atoms as a receiving antenna, Deb et al. directly recovered signal
at the baseband without any demodulation means \cite{a206}.
Anderson et al.
demonstrated an atomic receiver for amplitude modulation (AM) and frequency
modulation (FM) communication with a 3 dB bandwidth in the baseband of 100
kHz \cite{a222}. All of their work show atom-based quantum techniques are
new and promising candidates for microwave communication applications.

An RF E-field applied to the Rydberg atoms results in the Autler-Townes (AT)
splitting \cite{a227} of a ladder-type Rydberg electromagnetic-induced
transparency (EIT) transmission spectrum \cite{a228}. The measurement of the
RF E-field with Rydberg atoms can be converted into frequency measurement.
The AT splitting ($\Delta f$) is proportional to the RF E-field strength $E$
as described as \cite{a190}
\begin{equation}
2\pi \Delta f=\Omega _{RF}=\frac{\mu _{d}E}{\hbar },  \label{eq1}
\end{equation}
and inversely the RF E-field strength is expressed as
\begin{equation}
E=\frac{2\pi \hbar }{\mu _{d}}\Delta f,  \label{eq2}
\end{equation}
which quantitatively associates the RF E-field strength and the AT spectral splitting, where $\Omega _{RF}$ is the Rabi frequency of the Rydberg state transition
induced by the RF E-field and $\mu _{d}$ is the transition dipole moment of
the adjacent Rydberg states and can be
calculated accurately, and $\hbar $ is Planck's constant. We can see that a larger dipole moment $\mu_d$ corresponds a larger gain coefficient to magnify a weak electric field observable by the optical AT-splitting.
If the probe laser is scanned, the wavelength mismatch effect has to be included and the formula
above is corrected as \cite{a194, a180}
\begin{equation}
E=\frac{\lambda _{p}}{\lambda _{c}}\frac{2\pi \hbar }{\mu _{d}}\Delta f.
\label{eq3}
\end{equation}%
When we apply amplitude modulation (AM) to the RF carrier, probe
transmission also carries a relating modulation signal and can be directly
measured using a fast photo-diode detector. Owing to a potential high
sensitivity, the quantum receiver can be used for very weak signal
communication, which can also greatly reduce the cost of a transceiver
system \cite{a224, a214}.

To improve the sensitivity of MW sensing, a conceptually new microwave
electric field sensor called Rydberg-atom superheterodyne receiver
(Superhet) is demonstrated \cite{a188}. In an atomic superhet, a strong
on-resonant local MW field results in two dressed states $|\pm\rangle$
energetically separated by $\hbar \Omega_{L}$. They are, respectively, the
symmetric and antisymmetric superpositions of two bare Rydberg states. At
this critically separated point, a small MW signal will lead to a highest
dynamic optical response, which is used as the MW sensing.
In order to obtain electric field measurements with high accuracy, in their
work, they apply a local MW electric field ($E_{L}$) to achieve the
resonant transmission point $\overline{P}_{0}$ that has considerable slope $%
|\kappa_{0}|$. The local MW electric field is chosen as $E_{L}=$ 3.0 $\rm{mV \cdot cm^{-1}}$.
The sensitivity of this technique scales favorably, even up to 55 ${\rm nV
cm^{-1}\cdot Hz^{-1/2}}$ in a modest setup.

Principally, it is possible to further improve the sensitivity of electric
field measurements via optimizing the probe beam and coupling beam power
\cite{a194, a185}, local electric field strength and frequency detuning \cite
{a192, a211} together. It is also vital to evaluate the individual
contribution of each physical parameter in enhancing the MW detection
sensitivity. In this work, we study all the parameter optimizations as well
as the resonant transmission point of the local electric field, which is
expected to give a better sensitivity.

\section{experimental}

\begin{figure}[hbtp]
\includegraphics[width=3.3in]{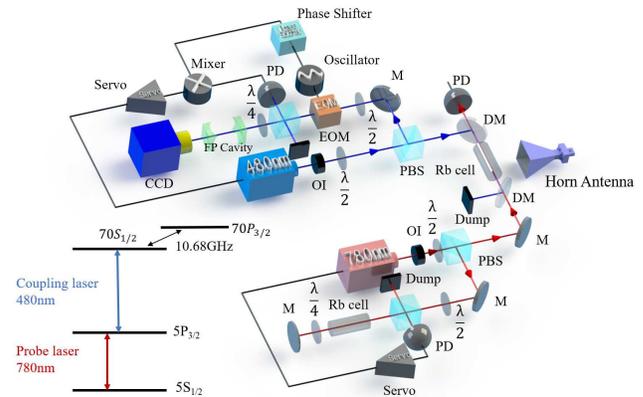}
\caption{Experimental setup and its concerned four-energy-level diagram. The
coupling ($\lambda_{c}=480$ nm) and probe ($\lambda_{p}=780$
nm) beams counter-propagate through a Rb vapor cell, forming a ladder-type
EIT with upper Rydberg states coupled further by microwave RF.
It is emit by a horn
antenna, serving  as a local RF electric field oscillator (LO). The probe beam passing through a Rb cell and a
dichroic mirror (DM) is detected by a photodiode (PD). The probe beam is
frequency-locked to $\rm{|5^2S_{1/2},F=3\rangle\rightarrow|5^2P_{3/2},F=3\rangle}$
transition via an integrated saturated absorption spectroscopy (SAS) unit of
$\rm{^{85}Rb}$ atoms. The coupling beam scans across transition $\rm{|5^2P_{3/2}\rangle%
\rightarrow|70^2S_{1/2}\rangle}$ and can also be frequency-locked to this
transition via a Pound-Drever-Hall technique (PDH) based on an ultra-stable
cavity on demand. }
\label{fig1}
\end{figure}

A typical schematic of our experimental apparatus is shown in Fig.\ref{fig1} with the related
energy levels of $\rm{^{85}Rb}$ in the inset \cite{a194, a185,a192, a211}. Theoretically two different laser
systems are needed to address the Rydberg levels in Rb, one near 780 nm and
one near 480 nm. The 780 nm light is tuned to the $D_2$ transition in Rb
commonly, corresponding to the transition from $\rm{|5^2S_{1/2},F=3\rangle%
\rightarrow|5^2P_{3/2},F=3\rangle}$, and the 480 nm one continues to excite the
atom on $\rm{|5^2P_{3/2}\rangle}$ to high Rydberg state $\rm{|70^2S_{1/2}\rangle}$,
forming a cascading irradiation configuration. The microwave RF couples the
Rydberg state to its neighbor partner $\rm{|70^2P_{3/2}\rangle}$ with its electric
dipole interaction strength monitored by the transparency of the probe beam
of 780 nm through the atomic vapor.

To avoid the Doppler background in the probe transparent spectrum, we record
the spectrum by scanning the coupling laser frequency of 480 nm while the
probe laser of 780 nm locked to a saturation absorption spectroscopy (SAS)
of Rb. The probe beam is derived from a cat-eye Morglab laser at 780.246 nm
with maximum output of 180 mW and a free running linewidth of $\sim1$
MHz. The coupling laser (480 nm) is generated by a frequency-doubled
amplified diode laser system (TA-SHGpro, Toptica, Munich, Germany), which is
scanned across the $\rm|5P_{3/2}\rangle\rightarrow|70S_{1/2}\rangle$ hyperfine
transition and real-time monitored by a FP-cavity for relative frequency
reference. Both lasers can  also be selectively locked to an ultralow
expansion glass (ULE) Fabry-Perot (FP) cavity using the Pound-Drever-Hall
technique (PDH) with two pairs of mirrors mounted on the same supporting
skeleton immune to the room temperature fluctuation, within less than 1 kHz.
The standard PDH-locking skills have been used in the feedback loop where
the laser beams imposed to the FP-cavity are modulated by an electro-optical
modulator (EOM) driven by a local oscillator of $\sim20$ MHz. The
locking signal comes from the mixing between the phase-shifted local
oscillator and the photo-diode signal of laser reflected from the FP cavity,
respectively. The 780 nm beam is focused to a spot of waist 750 $\rm{\mu m}$
with typical power of 100 $\rm{\mu W}$, corresponding to Rabi frequency
$\Omega_{p}=2\pi\times 11.13$ MHz. While the 480 nm beam focused to 1.25 mm
with  typical power of 420 mW, corresponding to Rabi frequency $\Omega_{c}=2\pi\times 0.69$ MHz.

Based on the optical EIT scheme that the coupling ($\lambda_{c}=480$ nm) and
probe ($\lambda_{p}=780$ nm) beams counter-propagate through a Rb vapor cell
of length 70 mm and diameter 25 mm, a horn antenna has been used to generate
a local RF E-field oscillator (LO) to couple the upper two Rydberg states
${\rm |70^2S_{1/2}\rangle}$ and $\rm{ |70^2P_{3/2}\rangle}$
with frequency $\sim10.68$ GHz via a horn antenna in free-space.
The transition dipole moment is calculated from the ARC library
\cite{92776} and its value is determined as $\mu_d=2395$ a.u..
In addition, a small signal (SIG) produced by a RIGOL function
waveform generator has already been amplitude-modulated to the RF LO to simulate the detection signal. The optimization experiment can be performed on this one-horn configuration.

\section{RESULTS}

\begin{figure}[hbtp]
\includegraphics[width=3.3in]{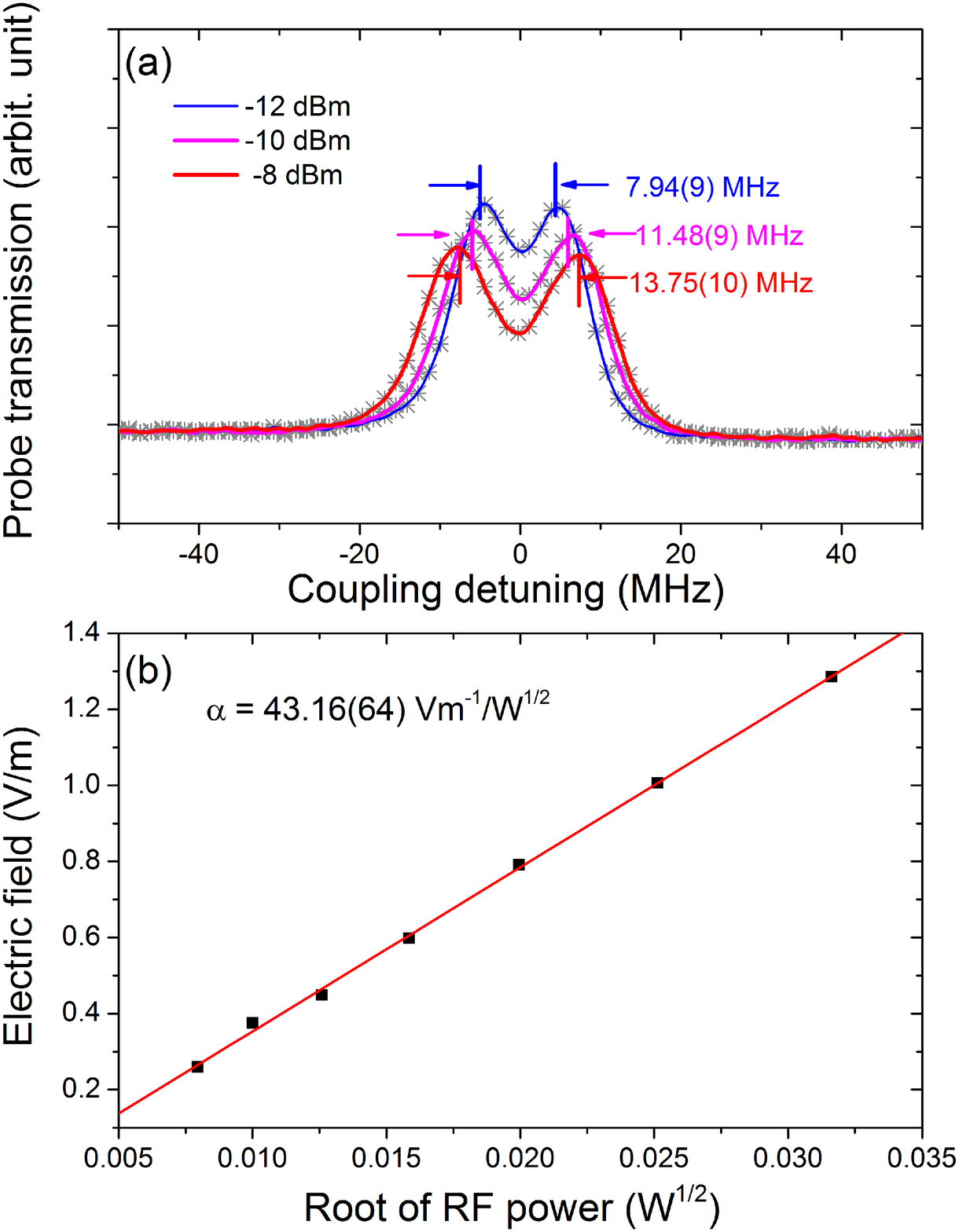}
\caption{(a) Typical AT-splittings due to the RF induced interaction
between the adjacent Rydberg states at different RF powers. (b) Measurements of the local RF
E-field $E_L$ (black square) by EIT-AT spectra versus the square root of the output
power, $\sqrt{P_{RF}}$, and its linear fit (red line) by formula $E_{L}=\alpha\sqrt{P_{RF}}$ with parameter determined as $\alpha=43.16(64)\rm{Vm^{-1}/W^{1/2}}$.}
\label{fig2}
\end{figure}

As we have mentioned previously, in the technique of superheterodyne,   a strong local MW has been  employed to induce considerable AT-splitting where the antennas has a highest dynamic response to the target signal MW carrier \cite{a188}.
Typical AT-splittings at various RF power and one of their dynamic responses are
shown in Fig.\ref{fig2}(a). The RF E-field couples a pair of specific Rydberg states,
$\rm{|70^2S_{1/2}\rangle}$ and $\rm|70^2P_{3/2}\rangle$. The typical AT-splittings are measured
at RF power -12, -10 and -8 dB, corresponding to AT-spectral line separation
of 7.94(9), 11.48(9) and 13.75(10) MHz.
We can deduce the electric field strength of the local RF field according to  Eq.\ref{eq3}.
Thus we can measure a series of data points of  RF E-field versus the applied RF power in the vicinity of a given LO power, for example, at the point corresponding to E-field of $3.53(18)\ \rm{mV\cdot cm^{-1}}$. The data are  shown in black
squares  in the Fig.\ref{fig2}(b).

As the RF E-field is proportional to the square
root of the output power,  given by the standard antenna equation
$E_{L}=\sqrt{30P_{RF}g}/d$ \cite{a188},
where $P_{RF}$ is the power of the microwave source, $g$ is a gain of the
antenna, and $d$ is a distance from the antenna to the cell. Considering $g$ and $d$ are constants for a certain  experiment, the relation can be simplified as
$E_{L}=\alpha\sqrt{P_{RF}}$, where $\alpha=\sqrt{30g}/d$ is the effective
gain of horn antenna. In our experiment, the linear least square fit determines
the effective gain as $\alpha=43.16(64)$ $\rm{Vm^{-1}\cdot W^{-1/2}}$.
The fit line is plotted in red solid line in Fig.\ref{fig2}(b).
The quality of the linearity of the
measurements is also very good, excluding the complex state interaction induced nonlinearity \cite{92738}. Therefore, the weak RF E-field value can be obtained
by the above formula at given antenna power.

\begin{figure}[hbtp]
\includegraphics[width=3.3in]{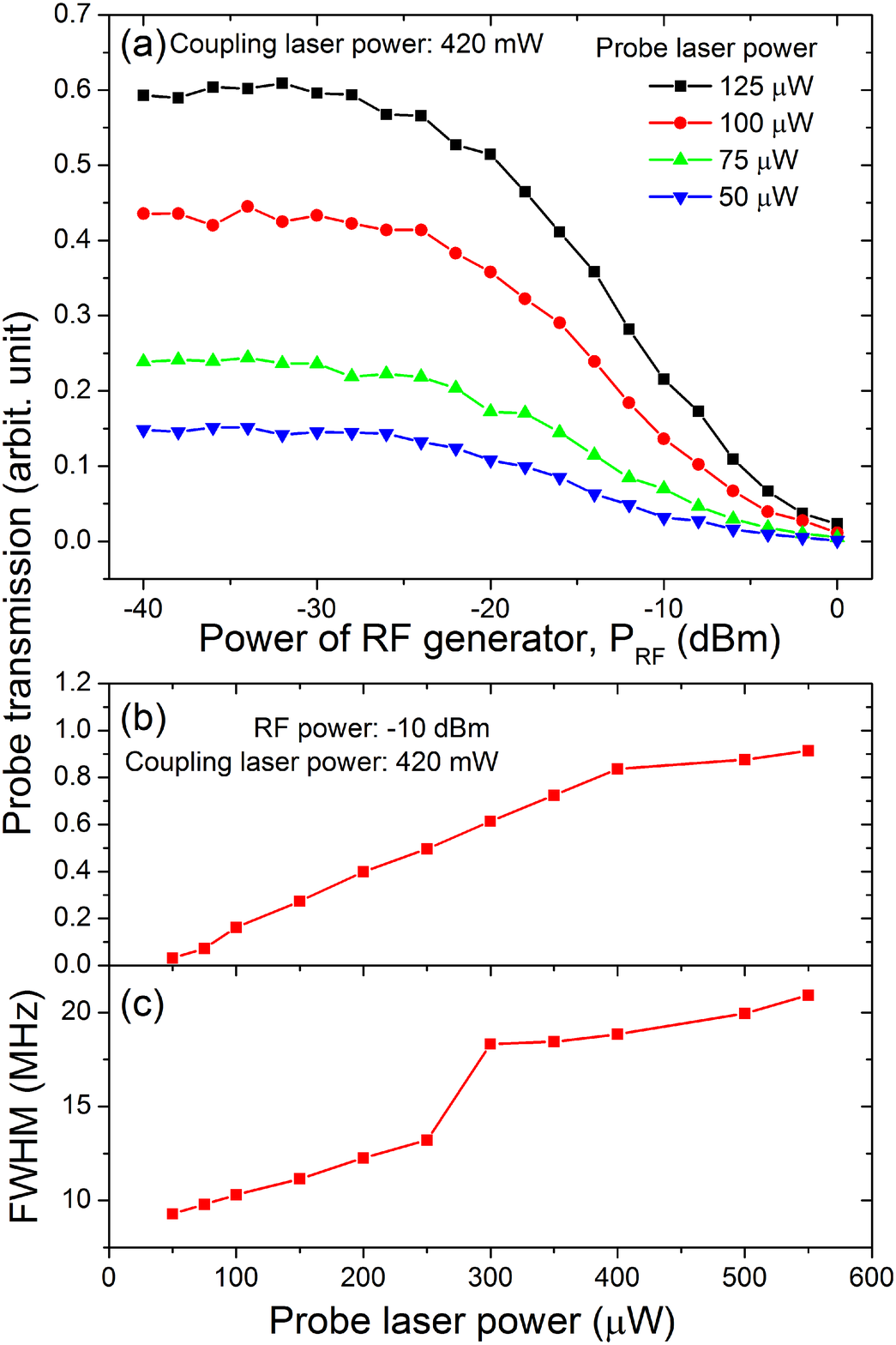}
\caption{(a) Probe optical transmission response on the applied RF power at various probe laser powers. (b) At the optimized RF power
of -10 dBm and coupling laser intensity of 420 $\rm{mW}$, the optical signal increases linearly with the probe laser intensity, but (c) the AT-spectral linewidth also almost linearly increases. An optimized probe laser power should keep a balance between them.}
\label{fig3}
\end{figure}

The probe optical transmission response is not linearly dependent on MW RF powers and the LO electric field strength should be chosen at the point with maximum slope \cite{a188}. As we know, at lower RF electric field, the AT-splitting is very small and the probe optical response to the applied MW power is not sensitive \cite{92717}. On the contrary, at very high RF carrier power, the AT-splitting is large enough where the splitting lines have already been separated. At this moment, the probe optical response is not sensitive, either. An intermediate case is preferred at which point the optical response has the largest slope on the applied RF power.
This point also relies on the power of probe and pumping laser powers. We present the measurement of the optical response on the power of RF generator in Fig.\ref{fig3}(a) at different probe laser powers.
We can see that a higher power of probe laser can bring larger optical transmission signal and also larger slope or response gain for RF power.
We choose the optimized RF power at -10 dBm (corresponding to an electric field of $3.53(18)\ \rm{mV cm^{-1}}$ at the vapor cell) and further investigate the optical response dependent on the probe laser intensity. It is shown in Fig.\ref{fig3}(b), where the response increases linearly with the probe laser intensity. However, unfortunately, the AT-spectral line width increases, either, nearly linearly. Specially, at power of 300 $\rm{\mu W}$, the probe laser is too strong, leading to a spectral splitting. We should balance these two factors in a real experimental operation. As can be seen later, we prefer to a value of 100 $\rm{\mu W}$ so as to keep narrower linewidth and to keep far away from  the critical point of splitting.

\begin{figure}[hbtp]
\includegraphics[width=3.3in]{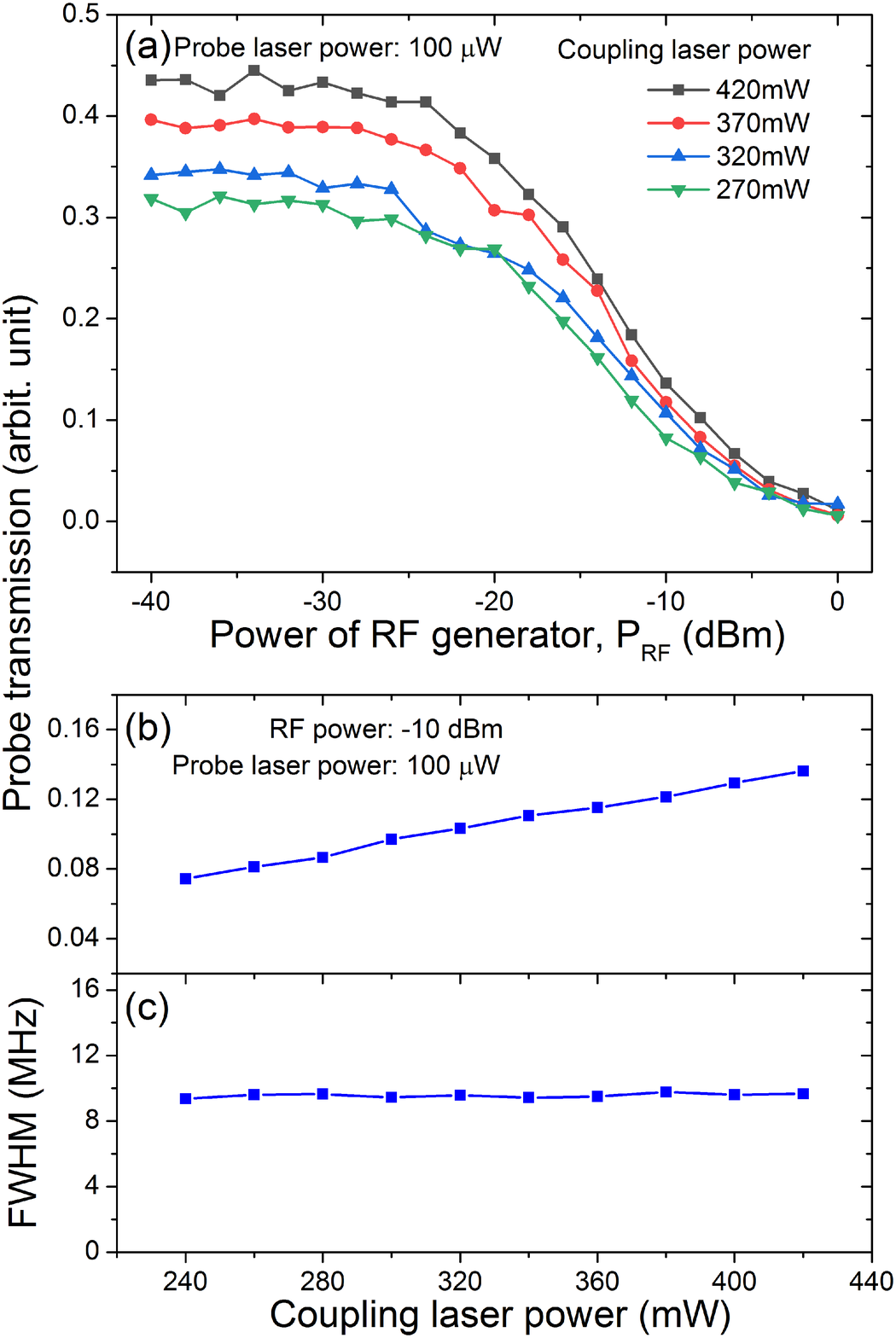}
\caption{(a) Probe optical transmission response on the applied RF power at various pumping laser powers. (b) At the optimized RF power
of -10 dBm and probe laser intensity of 100 $\rm{\mu W}$, the optical signal increases linearly with pumping laser intensity, while (c) the AT-spectral linewidth nearly unchanged.}
\label{fig4}
\end{figure}

The optimization also relies on the pumping laser power. We present the measurement of the optical response on the power of RF generator in Fig.\ref{fig4}(a) at different pumping laser powers.
Similar to the case of probe laser power optimization,  a higher power of laser can bring larger optical transmission signal and also larger  response gain for RF power.
At the optimized RF power of -10 dBm, the optical response dependent on the pumping laser intensity also shows a linear relation. It is shown in Fig.\ref{fig4}(b). Unlike the case of optimization of probe laser, however, the spectral linewidth keeps almost unchanged (Fig.\ref{fig4}(c)). It implies we can choose a higher coupling laser power if possible. It's also physically reasonable since a stronger coupling of 480 nm laser is more favorable  for the transfer of RF AT-splitting information down to the probe laser.

\begin{figure}[hbtp]
\includegraphics[width=3.3in]{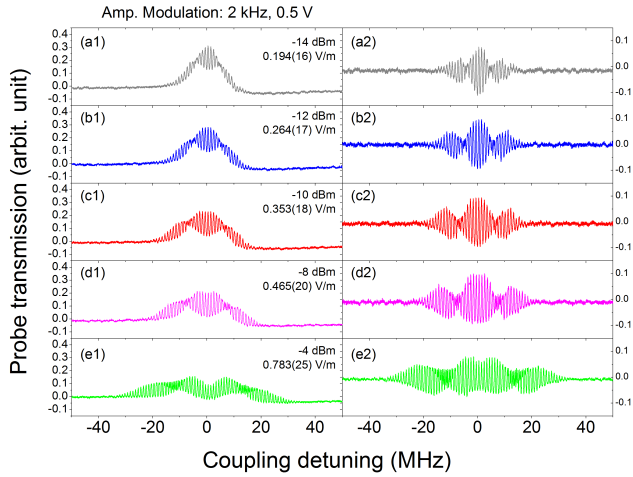}
\caption{The measurement of optical response of MW at different RF powers varying from $-14$ to $-4$ dBm (a1-e1) and the dynamical signal extraction (a2-e2). The laser powers are fixed at the optimized intensities. When the RF power increases up -10 or -8 dBm, the response has a large amplitude and nearly gets saturated, implying an optimized RF power.}
\label{fig5}
\end{figure}

We can also turn back to the optimization of the  MW RF power according to the dynamical response by applying a weak
amplitude modulation on the RF generator.
The measurement of optical response of MW at different RF powers varying from $-16$ to $-8$ dBm
is shown in Fig.\ref{fig5}(a1-e1). To see the dynamical response more clear, we extract the dynamical signal by filtering the flat DC part away, which is shown in Fig.\ref{fig5}(a2-e2).
We can see that when the RF power increases up -10 dBm, the response has a large amplitude and nearly gets saturated, implying an optimized RF power. The optimized RF power corresponds an electric field intensity of 0.353(18) V/m, close to the value of 0.3 V/m in Ref.\cite{a188}.
The above measurement is performed at laser power fixed at the optimized intensity in previous steps shown in Fig.\ref{fig3} and Fig.\ref{fig4}.
The amplitude modulation frequency is taken as 1 kHz with magnitude of 0.5 V at the input interface of the RF microwave generator, corresponding to a maximum electric field of 1.720(25) $\rm mV\cdot cm^{-1}$ at the vapor cell.

\begin{figure}[hbtp]
\includegraphics[width=3.3in]{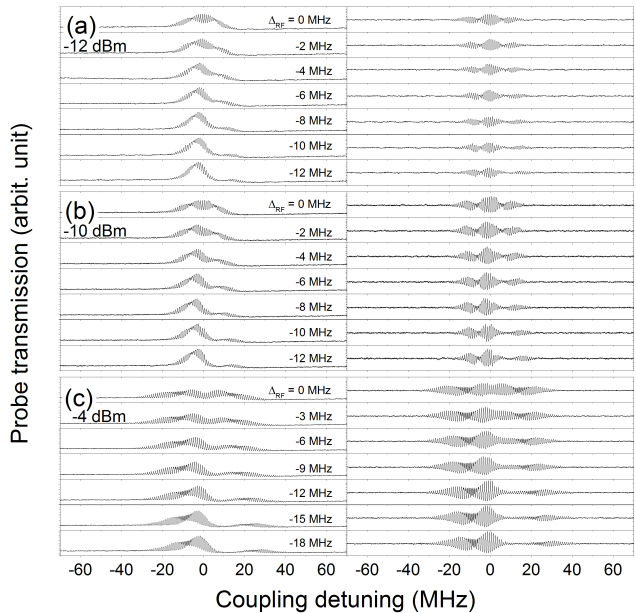}
\caption{
The measurement of optical response of MW at different RF frequency detunings (left) and its  dynamical signal extraction (right). A moderate detuning can increase the dynamical signal.
The laser powers are fixed at the optimized intensities. }
\label{fig6}
\end{figure}

To further improve the sensitivity of atomic antenna in MW communication, we detune the carrier MW frequency off resonance forming an asymmetrically splitting optical response, which is similar to the case of Cs atom \cite{a192}.  There the MW is detuned from the resonant transition between two Rydberg states, $\rm{47D_{5/2}\leftrightarrow48P_{3/2}}$, corresponding to a value of 6.946 GHz.
At this moment, we can compare the optical response of amplitude-modulated signal MW RF at a given
local RF power but at resonance and off-resonance to the transition between the
adjacent Rydberg levels. It is shown in Fig.\ref{fig6}.
As the same as in Fig.\ref{fig5}, the amplitude modulation frequency is taken as 1 kHz with magnitude of 0.5 V at the input interface of the microwave generator.
An intuitive view of Fig.\ref{fig6} tells us that a moderate RF detuning can help to increase the dynamical signal, especially at stronger RF power -10 or -8 dBm as shown in Fig.\ref{fig6}(b) and -4 dBm in Fig.\ref{fig6}(c).

\begin{figure}[hbtp]
\includegraphics[width=3.3in]{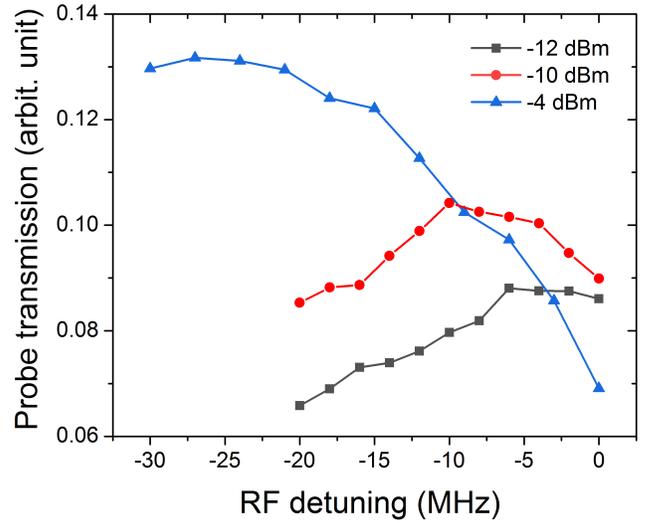}
\caption{The maximum dynamical signal response dependent on the RF detunings
at different MW RF powers in Fig.\ref{fig6}, where obvious optimization can be determined.}
\label{fig7}
\end{figure}

For better view of the result in Fig.\ref{fig6}, we can extract the maximum dynamical signal amplitude and redraw it in Fig.\ref{fig7}. We can see that
at RF power -12 dBm, the optimization is located at detuning $\Delta_{RF}=-6$ MHz but at higher RF power, -10 dBm and -4 dBm, the better LO power goes up to larger detuning values, -10 and -27 MHz, respectively. Specially, at RF power of -10 dBm, the dynamical signal is not sensitive to the RF detuning in a more wide range, -20 to -6 MHz. This feature is more useful in real experimental application.

\begin{figure}[btp]
\includegraphics[width=3.3in]{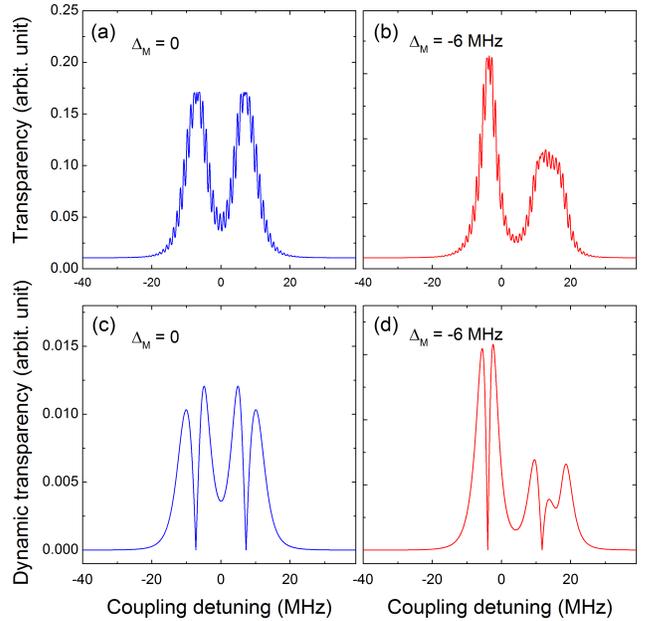}
\caption{Theoretical simulation of the dynamical optical response for the RF detuning. (a) and (b) are the probe transparency signal under the 1 kHz modulation at zero and -6 MHz RF detuning, respectively. Their dynamic optical response extractions are shown in (c) and (d), where a little increase in signal can be obtained by RF detuning.}
\label{fig8}
\end{figure}

The enhancement of  dynamic optical response by RF detuning can be understood by a theoretical simulation based on the optical Bloch equation considering the four energy levels \cite{92735,91262,92167,92738,89261,a188}.
 Taking the energy level of the intermediate state $\rm 5P_{3/2}$ as a reference, the Hamiltonian $H$ can be expressed as a matrix form
\begin{equation}
H= \left[
\begin{array}
[c]{cccc}%
\Delta_{c}+\Delta_{RF} & -\frac{\Omega_{M}}{2} & 0 & 0\\
-\frac{\Omega_{M}}{2} & \Delta_{c} & -\frac{\Omega_{c}}{2} & 0\\
0 & -\frac{\Omega_{c}}{2} & 0 & -\frac{\Omega_{p}}{2}\\
0 & 0 & -\frac{\Omega_{p}}{2} & -\Delta_{p}%
\end{array}
\right],  \label{eq2}%
\end{equation}
where a  probe laser
 couples the ground state $\rm5S_{1/2}$  and intermediate state $\rm5P_{3/2}$,
 characterized by  its
 Rabi frequency $\Omega_p$
while a coupling laser $\Omega_c$ induces an  interaction between  the intermediate state $\rm5P_{3/2}$ and the Rydberg state
$\rm70S_{1/2}$. $\Delta_i$ ($i=\rm p,c,RF$) depicts the frequency detunings of the probe, coupling beams and the microwave, respectively. Different from the usual Ladder-type EIT configuration, a microwave field is additionally applied and
correlates this Rydberg state to another close neighbor Rybderg state $\rm70P_{3/2}$ with large Rabi frequency $\Omega_{\rm RF}$ due to the considerable dipole moment between Rydberg states.
The simulation is shown in Fig.\ref{fig8}.
Theoretical simulations of the  optical transparency response for the RF detuning are presented in Fig.\ref{fig8}(a)-(b), corresponding to RF detuning of $\Delta_{\rm RF}=0$ and $-6$ MHz. A small amplitude-modulation has been applied to the RF field  under the 1 kHz modulation.
From the above spectra, we can obtain the corresponding  dynamic optical responses as Fig.\ref{fig8}(c)-(d). We can see that the strong dynamical response locates at the shoulder of the transparency spectral line rather than the spectral peak itself. Moreover, a little enhancement of the dynamical response can be achieved by RF detuning of -6 MHz at Fig.\ref{fig8}(d). The RF detuning causes the break-down of the symmetry of the optical response and transfers the dynamic optical gain to the branch closer to the zero detuning.

\begin{figure}[btp]
\includegraphics[width=3.3in]{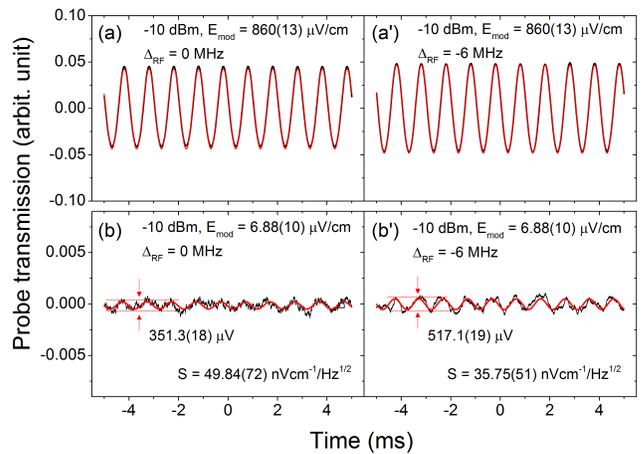}
\caption{The optical dynamical response at two RF detuning frequencies but all other parameters
fixed at the optimized values. The applied amplitude modulation intensity decreases until approaching the  lowest visual perceptible threshold.}
\label{fig9}
\end{figure}

Finally, we quantitatively evaluate the sensitivity of the Rydberg RF sensing system.
The applied amplitude modulation intensity decreases until approaching the  lowest visual perceptible threshold at RF frequency resonant to and detuned away from the transition between the two adjacent Rydberg states.
The optical dynamical response these two frequencies are shown in Fig.\ref{fig9} where all other parameters
fixed at the optimized values. The LO RF power is set to -10 dBm as previously determined. Fig.\ref{fig9}(a)-(b) correspond to the case of RF resonance while Fig.\ref{fig9}(a') and (b') to the case at the optimized detuning of -6 MHz. The large modulation gives a nice sinusoidal fit for the measured data points as presented in Fig.\ref{fig9}(a) and (a'), corresponding to a signal RF of 860(13) $\mu$V/cm.

When the modulation deceases to a lower perceptible threshold, corresponding to 6.88(10) $\rm{\mu V/cm}$ at the vapor cell, the detuning method remarkably has a better sensitivity, indicating a sensitivity improvement.
The latter optical response gives a high value of $517.1(19)\ \rm{\mu V}$ on the photo detector, larger than that of $351.3(18)\ \rm{\mu V}$ at zero detuning.
The method without detuning gives a sensitivity of 49.84(72) $\rm{nVcm^{-1}\cdot Hz^{-1/2}}$ while the detuning of -6 MHz gives a sensitivity of 35.75(51) $\rm{nVcm^{-1}\cdot Hz^{-1/2}}$.
A further step is to measure the noise spectrum with zero and -6 MHz detuning. The RF detuning of -6 MHz provides a sensitivity of 12.50(04) $\rm{nVcm^{-1}\cdot Hz^{-1/2}}$, better than that 13.69(40) $\rm{nVcm^{-1}\cdot Hz^{-1/2}}$ at zero detuning as well.
It corresponds a smallest discernible RF electric field of 176.78(57) $\rm{pVcm^{-1}}$, measured at time scale of 5000 s, better than that $780\ \rm pVcm^{-1}$ with sensitivity of
$55\ \rm nVcm^{-1}\cdot Hz^{-1/2}$ in Ref.\cite{a188}.

\section{Conclusion}
We have built up the highly excited Rydberg atom based RF field sensing system via electromagnetically induced transparency
with two color cascading lasers.
After choosing a higher Rydberg state as the sensing medium, we study the optical response of the probe beam dependent on the probe and coupling beam laser powers and fix them at the  optimized values. A further optimization is performed on the LO RF power where the AT-splitting is the most sensitive to the dynamic RF electric field. At the same time, we detune the RF field frequency to break the symmetry of the AT-splitting. It leads to one of the asymmetric branches more sensitive to the RF electric field variation, which further enhances the RF field sensitivity.
Based on the Bloch equation with four  energy levels concerned, we theoretically simulate the dynamic optical response of the system, completely supporting the optimization mechanism.

\begin{acknowledgments}
 This work is supported by National Natural Science Foundation of China (NSFC) (No. 12074388 and  No. 12004393).
\end{acknowledgments}

\end{document}